# Efficient Big Text Data Clustering Algorithms using Hadoop and Spark

Sergios Gerakidis
Hellenic Open University
Aristotelous 18, 26335
Patras, Greece

Sofia Megarchioti
University of West Attica
Agiou Spyridonos 28, 12243
Athens, Greece

Basilis Mamalis
University of West Attica
Agiou Spyridonos, 28, 12243
Athens, Greece

## ABSTRACT
Document clustering is a traditional, efficient and yet quite effective, text mining technique when we need to get a better insight of the documents of a collection that could be grouped together. The K-Means algorithm and the Hierarchical Agglomerative Clustering (HAC) algorithm are two of the most known and commonly used clustering algorithms; the former due to its low time cost and the latter due to its accuracy. However, even the use of K-Means in text clustering over large-scale collections can lead to unacceptable time costs. In this paper we first address some of the most valuable approaches for document clustering over such 'big data' (large-scale) collections. We then present two very promising alternatives: (a) a variation of an existing K-Means-based fast clustering technique (known as BigKClustering - BKC) so that it can be applied in document clustering, and (b) a hybrid clustering approach based on a customized version of the Buckshot algorithm, which first applies a hierarchical clustering procedure on a sample of the input dataset and then it uses the results as the initial centers for a K-Means based assignment of the rest of the documents, with very few iterations. We also give highly efficient adaptations of the proposed techniques in the MapReduce model which are then experimentally tested using Apache Hadoop and Spark over a real cluster environment. As it comes out of the experiments, they both lead to acceptable clustering quality as well as to significant time improvements (compared to K-Means - especially the Buckshot-based algorithm), thus constituting very promising alternatives for big document collections.

## 1. INTRODUCTION
Nowadays, data are being collected at huge speeds and spread everywhere around us; various news agencies produce thousands of news articles while Twitter generates over 500 million Tweets every day. Text management and retrieval methods over big documents data get progressively of great importance. Document clustering [1] is known as a highly efficient, machine-learning technique when a better insight of the documents of a collection is needed. K-Means is one of the most commonly adopted methods of clustering, due to its high quality results and low time cost. However, using the K-Means algorithm in document clustering may lead to unacceptable time overheads, since the cost of a K-Means iteration tends to grow as the number of iterations grows. The need for effective text clustering techniques (with respect to both efficiency and quality, and especially over big text collections) has triggered the investigation of probable modifications on an already effective technique like K-Means.

At the high level, many clustering techniques follow the specific procedure given below: after the proper initialization some iterative process is executed until the proper convergence criteria is met. In every iteration, the cluster membership is adjusted/updated for each one of the data points. So, in order to speed up such clustering methods, there are three main alternatives [2]: (a) by reducing the number of iterations (like the one-pass algorithms – CURE [3], BIRCH [4] etc.), (b) by restricting the access to the data points (e.g., by sampling / randomized sampling methods [5-8] – CLARANS etc.), and (c) by parallelizing/distributing the calculations [9-10]. Especially with respect to the latter category, the most recent and valuable research works are based on modern massive distributed memory processing frameworks like Hadoop and Spark. There are many such parallel/distributed efficient implementations based either on K-Means or on the hierarchical agglomerative clustering algorithm, as well as sampling algorithms that intend to improve clustering in both execution time and accuracy.

In [11], the authors, having as their main objective to improve K-Means execution time, first describe a relatively time intensive procedure to compute the optimal initial centers, which however is subject to very efficient parallelization in MapReduce; thus providing satisfactory total execution times over big text collections, especially when multiple nodes are used. In [12], an efficient Bisecting K-Means approach is presented, which is experimentally proved to lead to better results in text clustering than the typical K-Means approach. First, the typical K-means method is used to produce two intermediate clusters (sub-clusters) from the initial dataset. The sub-cluster with the lowest similarity is chosen to replace the input dataset. This procedure is repeated appropriately until the proper final number of clusters is reached.

In [13], a parallel hierarchical clustering approach (using the MapReduce framework) based on random partitioning is proposed to improve the time performance of hierarchical clustering. Specifically, the idea of dendrogram alignment is introduced, describing a technique to merge the dendrograms formed locally to construct a global one. The proposed method provides good execution times as well as high scalability. The implementation of our Buckshot-based adaptation presented in this paper, follows in some extent the techniques used in the implementation of the hierarchical clustering approach in [13].

In [14], a more theoretical distributed approach (based on the single-link hierarchical clustering method) is presented using MapReduce (DiSC). The key idea is to adjust the problem of clustering to the problem of finding the minimum spanning tree (MST) in a complete graph induced by the input set of data. Following the divide-and-conquer method, the initial set of data is partitioned into s splits and each couple of these splits construct a subgraph. The well known Prim's algorithm is then used to compute the MSTs locally for each subgraph. The well known Kruskal's algorithm is also used to merge the local MSTs for every K subgraphs, until all nodes become members of the same MST. Note also that the above methods [13,14] have been tested experimentally over real cluster environments and they have achieved quite satisfactory response times over big input sets of input data, especially when multiple nodes are used. The clustering quality is also kept at high levels due to the positive effect of the hierarchical clustering procedure. Further, in [15], an efficient MapReduce technique is presented that computes the connected components in logarithmic number of iterations for large graphs. Four different hashing schemes are given, and one of them has been proved to finish in $O(\log n)$ iterations, achieving $O(k(|V|+|E|))$ communication cost at round k.

Several other valuable research works and implementations over MapReduce can also be found in [16-20]. The use of sampling-based techniques as well as the possibility of applying one algorithm on a sample of the initial dataset and completing the clustering using another algorithm, usually form the basis for corresponding efficient implementations. In an analogous manner, the use of Spark framework for even faster related implementations has been addressed adequately the last years. Such a notably efficient KMeans-based is demonstrated in [21], whereas in [22] a highly efficient parallelization of the hierarchical agglomerative clustering method in Spark is also presented. A more detailed review on efficient parallel clustering algorithms for big data in Spark framework can be found in [29].

In this paper, having as our main objective the efficient handling of big text/document data, we first present a variation of an existing KMeans-based clustering algorithm, known as BigK-Clustering (BKC [23]), so that it can be applied in documents represented as weighted term vectors in the Vector Space. BKC clustering is a very effective clustering technique designed and evaluated for point data, and it's based on the idea of beginning with a big number of compact micro-clusters and keeping them inseparable and progressively growing till the end. It leads to clustering accuracy very close to the one of K-means, however in significantly less execution time, by reducing adequately the number of the necessary iterations. It may also be efficiently implemented in the MapReduce model, thus leading to a proper solution for big data clustering (see also [30]). Additionally, we present a combined method, investigating the potential of using two different algorithms in one, i.e. executing one algorithm on a sample of the initial set of data and completing the clustering procedure by applying another algorithm based on the preceding output. More specifically, we initially follow the idea of the Buckshot clustering algorithm given in [16], which first executes a hierarchical algorithm on a sample of the input set of data and then applies the results as initial centroids for a KMeans based routine with very small number of iterations. Furthermore, we develop an adapted version of the Buckshot approach suitable for (big) text data, and we finally present a highly efficient implementation of the proposed Buckshot-based adaptation in the MapReduce model (see also [31]).

Both the above promising approaches are then implemented using Hadoop and Spark frameworks to manage large-scale document collections efficiently, and they are experimentally tested over a real cluster platform and real text/document data of size up to 1GB. As it comes out of the experimental testing, they both exhibit acceptable performance with respect to clustering quality and accuracy, as well as substantial execution time improvements (up to 75-85% on average comparing to K-Means) and good speedup values (especially for 10 nodes). Moreover, a large number of experiments have also been performed over Spark, thus leading to significant additional improvements, especially when large amounts of input and intermediate data can be cached and processed iteratively in memory.

The rest of the paper is organized as follows. In Section 2 an overview of the parallelization efforts of K-Means in the MapReduce model is presented. In Section 3 the proposed variation of the BigKClustering algorithm is described as well as its implementation in MapReduce. In Section 4 the proposed Buckshot-based clustering approach is described as well as its suitable implementation in the MapReduce model. In Section 5 our extended experimental results are presented, and Section 6 concludes the paper.

## 2. K-MEANS OVER MAPREDUCE
A lot of research work has been attempted towards the proper massive parallelization of the K-Means method for text data (e.g. [24-26]). Most of them are based on the vector space model representation with tf-idf weights, and they usually lead to highly scalable implementations, providing significantly reduced response times when many distributed memory nodes are used. Moreover, in [27] a novel related method is described, in which the native K-Means algorithm is applied in combination with along the NMF factorization technique (Non-negative matrix factorization), thus resulting to notably more satisfactory results. Considering the most known implementations of K-Means in MapReduce in the literature, the one (PKMeans) given in [26] by Zhao et al may be regarded as one of the most efficient too. In the following, an overview of this approach is given, since it has also been used as the main basis for our BKC implementation introduced in the next section. In brief, the map function finds for each object of the dataset the closest center to be attached, while the reduce function computes the new centers. Also, a combiner is used to perform partial reductions within the same map task. The corresponding steps of the PKMeans implementation are described in more details below.

*Map function:* The initial set of data is stored as a sequence file of pairs in the <key-value> form. Every pair represents a data point. The key stands as the byte offset of the structure containing the data to the start of the sequence file, while the value is a string representing the data contained in the structure. The whole set of data is then divided into splits and it is passed to the mappers, where each data point is attached to the closest of the centers. All the necessary distance calculations are executed in parallel. Moreover, a global structure variable has to be created for each map task, which contains the clusters centers, so that each mapper can compute the closest center to every portion of its split.

*Combine-function:* After the completion of a map task, a combiner is used to combine the locally computed results of the same map task. The values of the points attached to the same center are then summed and the mean value is extracted.

*Reduce-function:* The output values of the combine function are given to the reducers. The partial sums of values that are labeled with the same key are also summed and a new mean is computed. This new mean value is considered as the new center of the cluster.

## 3. THE BIGKCLUSTERING APPROACH FOR DOCUMENTS

In the following the BigKClustering approach for efficient clustering of documents (instead of points) is presented, as well as its parallel implementation in the MapReduce model.

### 3.1 Adaptation for Documents Clustering

*Comparison measure.* In the original paper, the Euclidean distance is adopted as a metric to decide if a document should be assigned to a specific cluster. Here, we use the cosine similarity as the comparison measure, since it is more effective when we deal with high dimensional data such as documents in the vector space.

*Micro-cluster definition.* First, the notion of 'micro-cluster' (MC) has to be defined adequately, so that the similarity among documents (through the cosine similarity measure) can be reliably computed. In its typical definition, a component of the vector indicating a micro-cluster is kept for the longest distance met within the micro-cluster, that is between a point attached to the micro-cluster and the micro-cluster center. In our adaptation, however, we have to use the cosine similarity measure, so the 'longest distance' should be replaced with the 'lowest similarity' met within the micro-cluster. The adapted definition for a micro-cluster (in case of document collections) is formulated as follows: Having a group of d-dimensional vectors $X = (x_1, \ldots, x_n)$, a micro-cluster should be defined as a vector of $2d + 3$ dimensions in the next form: $(n_i, CF1_i, CF2_i, Center_i, min_i)$ where: (a) $i$ is the micro-cluster id, (b) $n_i, CF1_i, CF2_i$ refer to N, LS, SS in a CF vector, respectively, (c) $Center_i$ is the document we select from X to serve as the initial center of the micro-cluster, and (d) $min_i$ is the min cosine similarity noted between a document and a center, while the procedure for the assignment of documents to the initial centers takes place.

*Equivalence relation.* Also, the equivalence relation has to be redefined according to the cosine similarity measure. The appropriately adapted (for the case of documents) equivalence relation is formulated as follows: Having a set of micro-clusters MC, to any two micro-clusters $S_i, S_j \in MC$, if the similarity between them is greater than a threshold s (connection similarity), or there is a group of micro-clusters $(S_i, S_{t1}, S_{t2}, \ldots, S_j)$ among which the similarity between any two adjacent micro-clusters is greater than s, then we may define $S_i, S_j$ as having an equivalence relation. In our approach the similarity of two micro-clusters is:

$$sim(S_i, S_j) = cosSimilarity(Center_i, Center_j) - min_i - min_j$$

where the cosine similarity measure is considered between two documents. If the similarity between $S_i, S_j$ is not a positive value or zero, it is set to be zero. If the above similarity equals to zero and the similarity between $Center_i$ and $Center_j$ is larger than $min_i$ or $min_j$, $S_i, S_j$ may also be defined as having an equivalence relation. The set of micro-clusters having an equivalence relation is then connected to a group of micro-clusters.

### 3.2 The BigKClustering for documents

Like the original algorithm, our adaptation will attempt in each phase to keep the compactness of documents that are grouped together, which is subject to probable loss during the phase of constant regrouping. The main objective of the adapted BKC algorithm is to group closed to each other documents and form suitable groups of micro-clusters.
| *Algorithm* BigKClustering for documents $(ds, BigK, k)$ |
|---|
| Input: initial dataset $(ds)$, micro-clusters number $(BigK)$, final document clusters number $(k)$ |
| Output: clusters of documents |
| 1. Randomly select $BigK$ centers from $ds$.<br>2. Assign all documents in $ds$ to their most similar centers. Thus $ds$ is divided into BigK portions $p_1, \ldots, p_{BigK}$.<br>3. Form $BigK$ micro-clusters $s_1, s_2, \ldots, s_{BigK}$. Every micro-cluster corresponds to one data portion.<br>4. Assign an initial value to the connection similarity $s$ by taking the mean of all the min values.<br>5. Calculate the groups of micro-clusters by running the procedure joinToGroups $(\langle s_1, s_2, \ldots, s_{BigK}\rangle, k, s)$<br>6. Compute the centers of all the groups and let them to be the centers of the final clusters.<br>7. Attach every document in $ds$ to its most similar center and give it the proper cluster label. |
| *Procedure* joinToGroups $(\langle s_1, s_2, \ldots, s_{BigK}\rangle, k, s)$ |
| Input: micro-clusters list $(\langle s_1, s_2, \ldots, s_{BigK}\rangle)$, final document clusters number $(k)$, connection similarity $(s)$.<br>Output: groups of micro-clusters. |
| 1. For $i = 1$ to $BigK$ do<br>    $j = i - 1; f = true;$<br>1.1 if $(j\ != 0)$<br>    For $k = 1$ to $j$ do<br>1.1.1 if $(sim(s_i, s_j) == 0)$<br>    if $(cosSimilarity(Center_i, Center_j) \geq min_i \parallel cosSimilarity(Center_i, Center_j) \geq min_j))$<br>    $f = false; break;$<br>1.1.2 else if $(similarity(s_i, s_j) \geq s)$<br>    $f = false; break;$<br>1.2 if $(f == true)$<br>    provide $s_i$ with a new group id;<br>1.3 else<br>    Attach $s_i$ to the same group as $s_k$;<br>2. If the number of groups is $k$, then proceed to step 4. If it's not the case, go to step 3.<br>3. Adapt $s$ and go to step 1.<br>4. Return the groups of micro-clusters. |

**Fig 1: The adapted BigKClustering for documents**

First, we select at random BigK initial centers from the original centers. Then, each document of the dataset is attcahed to the closest (most similar) center. The above step is like one iteration of the K-Means procedure. Then the micro-cluster vectors are constructed from the randomly selected centers and their attached documents. After the micro-clusters are built, we give an initial value to the connection similarity s; specifically the mean of all the min values defined in all micro-clusters. Next, the equivalence relation is applied to construct groups of micro-clusters. The connection similarity value is adjusted and the equivalence relation is applied iteratively, until the number of the groups becomes equal to the desired clusters number. Finally, one iteration of the KMeans procedure is applied as the concluding step of the algorithm, in which the centers of the groups selected above are considered as the initial centers of the final clusters. Our proposed adaptation of the BigKClustering algorithm is shown more clearly in Fig. 1. In the 1st part the main steps of the algorithm are summarized, whereas in the 2nd part the joinToGroups procedure is given separately, where the groups of micro-clusters are constructed.

### 3.3 The MapReduce implementation

In this paragraph, the way the adapted BigKClustering algorithm was implemented over the MapReduce model is presented. Three corresponding MapReduce tasks were necessary to complete our implementation efficiently, which are described in details below.

The first three steps (steps 1,2,3) were implemented using a single MapReduce job, with multiple mappers. However, only one reducer was needed. Each mapper is assigned an input split (from the initial dataset) and outputs pairs in the <key-value> form, where the key is the closest center and the value is a document attached to this center. The reducer then builds the micro-clusters based on the outputs from the mappers.

Steps 4,5,6 were implemented in a separate MapReduce job, using a single mapper and a single reducer. The mapper was set to calculate the initial connection similarity and pass the computed value to the reducer. The reducer then performs the task of formulating the groups of micro-clusters.

Finally, the implementation of step 7 was performed through a third MapReduce job with multiple mappers and reducers. In this job the final clustering takes place, based on a similar method like the one of the first job. However, here, the input of the mappers consists of the centers of the groups of micro-clusters constructed in the second job. The results extracted from the map phase are passed to the multiple reducers and the final clusters are built.

## 4. OUR BUCKSHOT CLUSTERING APPROACH

As discussed in section 1 the use of sampling based techniques can lead to substantially reduced response times in big text data clustering implementations. However, in such algorithms the necessity to keep acceptable accuracy and quality is also crucial. Following this direction, we've developed an adapted version of the Buckshot method, taking advantage of the hierarchical agglomerative clustering method to preserve accuracy and quality. The Buckshot clustering approach is an adequate hybrid clustering method, mainly based on the combination of hierarchical and partitioning clustering techniques. It has been designed appropriately to gain from the advantages of both the former and the latter (i.e. the accuracy of the hierarchical methods and the reduced execution time of the partitioning methods).

More concretely, the existence of some hierarchical algorithm is assumed in the beginning, which clusters well however it's quite slow) and it's initially applied on a sample of the document collection. The above procedure is known as 'the cluster subroutine'. In the proposed approach, we've adopted the single-link criterion [1,16,27] for this subroutine (instead of group-average, complete-link or other related forms), since our main objective is to obtain not very tight clusters with good quality. This choice has been regarded as the most adequate one considering the big extent of the input data. Consider also that despite of the fact that the hierarchical clustering technique is time-intensive, this it isn't a crucial disadvantage if it's applied over a restricted in size dataset (like a proper sample of a large collection).

---

Initial Conditions:

s = number of documents in the initial sample,

n = total number of documents in the collection,

k = the desired clusters number

Basic Steps:

1. Choose 's = $\sqrt{kn}$' documents randomly

2. Execute the proper subroutine (single-link hierarchical agglomerative clustering) for the initial clustering of the selected 's' documents:

a. Construct the similarity matrix D (which provides the pair-similarities of all the s documents)

   i. Attach each one of the documents di to cluster ci

   ii. For each pair of clusters (ci, cj) with (i≠j) compute the similarity $SIM_{ij}$ among ci and cj, by applying the cosine-similarity measure

b. Form the 'k' Clusters

   Iterate *n-k* times:

   i. Search for clusters i and j which exhibit the largest similarity $SIM_{ij}$ with respect to single-link definition

   ii. Replace clusters i,j by an agglomerated cluster h

   iii. Adjust D to incorporate the new similarities between h and each one of the other clusters

   Extract the 'k' initial clusters

3. Group the rest of the documents [attache them to the 'k' clusters extracted above]

a. Compute the centroids of the 'k' clusters

b. Iterate for each one ('d') of the (totally 'n-s') documents which have not been initially clustered:

   i. Compute similarity between d and each centroid ci

   ii. Attach *d* to cluster *i* where sim(d, ci) > sim(d, cj) | i≠j

c. Iterate 'a' and 'b' steps for a small number of times to adjust the cluster output

d. Extract the 'k' final clusters

**Fig 2: The Buckshot-based clustering algorithm**

### 4.1 Description of the Buckshot-based Approach

Considering the beginning step of the algorithm, a sample of s= $\sqrt{kn}$ documents is first selected at random from the document collection. The specific initial subroutine chosen above is applied as the high-precision clustering routine over the documents of the sample, to extract the initial centers (1st phase – steps 1 and 2 in Fig. 2). Next (2nd phase / step 3 in Fig. 2), the initial generated centers should be used as the main basis for clustering the whole collection, by attaching the rest of the documents to the closest (most similar) initial center. The typical Buckshot method does not imply something with respect to which should be the best choice for this task, although several alternatives are proposed. In this work we've followed an iterative K-means-based (assigning each document to the most similar center) algorithm with two or three iterations. A more clear view of our proposed adaptation of the Buckshot algorithm for big document collections is provided in Fig. 2. The execution of the Buckshot algorithm typically needs linear time (since s= $\sqrt{kn}$, the total time needed is equivalent to O(kn) where k has a much lower value than n), thus providing a quite fast solution. The latter guarantees the viability of this method for clustering large-scale document collections too. However, when very large or huge document collections are considered, the restriction of linear execution time makes the use of the algorithm inappropriate for several kinds of applications. On the contrary, it can be regarded as a very good solution if it's supposed to run over small or medium-size datasets. To overcome this inefficiency in any case, we've followed the direction of applying massive parallelism over a shared nothing (distributed memory) parallel environment, having as our main objective to gain efficient behavior even for very large / huge text data.

## 4.2 The MapReduce Implementation

In the following we describe the proposed implementation of our Buckshot-based approach in the MapReduce model. In the beginning, with respect to the parallelization of the setup phase of the random selection (step 1 in Fig. 2), we've adopted a direct parallelization procedure with many mappers assigning random integer keys to the input, and one reducer extracting the output.

### 4.2.1 The HAC-based module implementation

With respect to the parallelization of the first phase (step 2 in Fig. 2), we've suitably adjusted the approach given in [13], which can lead to significant improvements in the time cost of the hierarchical clustering algorithm. Following this approach a parallel HAC algorithm based on random partitioning (and using the MapReduce framework) is proposed to get relevant improvements in the time efficiency of the hierarchical clustering procedure. Moreover, the notion of dendrogram alignment is properly introduced here, which is an efficient mechanism to merge locally extracted dendrograms to construct a global one. The initial set of data is split at random in smaller partitions on the mappers. Each partition is passed to single reducer, where the sequential hierarchical clustering algorithm is executed. The extracted dendrograms by this local procedure are finally then aligned to each other using a suitable global dendrogram alignment procedure. The relevant mapper and reducer subtasks of the whole implementation are analyzed in more details below.

*Map function:* Supposing that the number of the initially chosen documents is equal to 's' (as shown in Fig. 2) and we have M partitions in total, the number of documents assigned in each partition is calculated as $np = s/M$ and it is passed as a relevant parameter to the mappers. Let $S_m$, $m = 1, 2, 3,…, M$ be the number of items/documents the $m^{th}$ partition may accept. Also suppose that the initial value of $S_m$ has been set to np. Every mapper gets a document one at a time and it generates a <key,value> pair. The data is considered as the value whereas a random integer i (from 1, 2, 3, ..., M), is considered as the key. The probability of the partition index m is proportional to $S_m$, which means that at the end the number of items in each partition is consistent with n and the workload of all the local clustering tasks is equivalent. Eventually, all the <key,value> pairs are collected and sent to the reducers, grouped by their keys.

*Reduce function:* In every reduce task, the reducer only works with the <key,value> pairs having the same key. More concretely, it applies the sequential hierarchical clustering technique on the data it has, and the dendrograms generated locally are stored for future use.

### 4.2.2. The K-means-based module implementation

With regard to the parallelization of the second phase (step 3 in Fig. 2) we've appropriately adapted a K-Means based implementation, and more concretely, we've followed the approach described in section 2. The latter was also the most preferable solution due to the need of a fair comparison with the BKC algorithm given in section 3. Briefly speaking, following this implementation in our case, the *map function* first assigns each object (document term vector) of the 'n-s' documents not initially clustered, to the closest center, while the *reduce function* updates the new centers. A combiner is also applied to partially combine the intermediate values during the same map task. Each document is supposed to have a <key-value> pair representation, as described in more details in section 3.

## 5. EXPERIMENTAL RESULTS

The proposed adjustment of BKC algorithm for clustering of documents - instead of points - as well as our proposed customization / adaptation of the Buckshot algorithm in the MapReduce model were initially implemented using Apache Hadoop and Java, and their performance was compared to the performance of the native K-means algorithm implemented in MapReduce too. The experimental evaluation was performed in a cluster of 10 nodes, each of which has a 3.0GHz quad-core processor, 4GB RAM, 500GB hard drive and 1Gbps network connection. For the purposes of the paper, the 20_newsgroups collection was adopted, which is a collection of about 20000 postings in 20 newsgroups and generates a file of vectors of 80.2MB. In order to measure the performance of our approaches in sufficiently large-scale data we've also generated a relevant synthetic collection of almost 1GB, by multiplying the original 20_newsgroups collection adequately. Corresponding measurements have also been taken with respect to the implementation of the proposed Buckshot-based approach in Spark framework. Here, the implementation was based on Scala language, which offers both simplicity and increased flexibility by effectively combining object oriented and functional programming features. We present the most indicative results in the next paragraphs.

## 5.1 Evaluation of the BigKClustering Approach

First, the results with regard to the evaluation of the adapted BKC algorithm are presented. Tables 1, 2 and 3 give the measurements of the construction of 50, 100 and 200 clusters respectively, with both algorithms, using the initial collection 20_newsgroups. In the case of BigKClustering, 50, 100 and 200 initial centroids are generated from 250, 300 and 450 micro-clusters respectively. After testing, it was observed that K-Means converges after 8 iterations. The execution time and the RSS value after the convergence of K-Means are presented, as well as the respective measurements for BKC after the completion of all three MapReduce tasks. The execution time of both algorithms was measured using 1, 3 and 10 nodes/reducers. Table 4 shows also the measurements of the construction of 400 clusters, generated from 800 micro-clusters using our the synthetic collection of 1GB. As it comes out from Tables 1, 2 and 3, the execution time of the BKC algorithm is significantly less than the one of K-Means in almost all cases. More concretely it leads to execution time improvements of up to 85% compared to the total time required for the convergence of K-Means. Especially when the bigger values of k are applied the time improvements are close to the above maximum for all the varying numbers of nodes (1, 3 and 10), thus demonstrating the high scalability of the proposed implementation. Moreover, the quality of clustering is kept in acceptable levels since the achieved RSS value is quite close to the final RSS value of K-Means (having a difference between 5% and 8% in all cases).

The above observations are also validated by the additional measurements presented for the synthetic collection of 1 GB in Table 4 (improvements of up to 78% can be observed).

Furthermore, in Table 4 the corresponding measurements with respect to the implementation of the proposed BKC approach in the Spark framework are also given. As it was expected (due to the in-memory iterative processing taking place in Spark) substantial additional improvements have been observed, leading to the capability of clustering 1GB of text data in only 10 minutes using the proposed BKC algorithm. The corresponding improvement comparing to KMeans remains also at notably high levels (i.e. it raises up to 70%). These spectacular achievements will be further improved as we'll see later on with the use of our hybrid Buckshot-based clustering approach.

**Table 1. BKC - 20_ngroups (n=20000), k=50, K=250**

| Nodes | Execution Time | | RSS | |
|---|---|---|---|---|
| | KMeans | BKC | KMeans | BKC |
| 1 | 77m57s | 10m7s | 140.39 | 148.52 |
| 3 | 43m25s | 5m44s | 140.12 | 148.42 |
| 10 | 11m41s | 3m7s | 140.78 | 150.19 |

**Table 2. BKC - 20_ngroups (n=20000), k=100, K=300**

| Nodes | Execution Time | | RSS | |
|---|---|---|---|---|
| | KMeans | BKC | KMeans | BKC |
| 1 | 81m55s | 12m57s | 135.82 | 144.87 |
| 3 | 54m30s | 7m36s | 135.19 | 145.30 |
| 10 | 15m55s | 3m 55s | 136.21 | 145.27 |

**Table 3. BKC - 20_ngroups (n=20000), k=200, K=450**

| Nodes | Execution Time | | RSS | |
|---|---|---|---|---|
| | KMeans | BKC | KMeans | BKC |
| 1 | 158m21s | 20min15s | 129.17 | 137.95 |
| 3 | 91m33s | 12min44s | 128.92 | 137.76 |
| 10 | 28m45s | 4min9s | 129.53 | 139.34 |

**Table 4. BKC - 1GB (n=250000), k=400, K=800**

| Nodes | Execution Time | | RSS | |
|---|---|---|---|---|
| | KMeans | BKC | KMeans | BKC |
| 10 (MR) | 324m41s | 71m41s | 1131.30 | 1195.36 |
| 10 (Spark) | 33m35s | 10m4s | 1118.30 | 1187.47 |

## 5.2 Evaluation of the Buckshot Approach

Second, the corresponding results with regard to the evaluation of the our Buckshot-based approach are presented. Tables 5, 6 and 7 give also here the measurements of the construction of 50, 100 and 200 clusters respectively, with both algorithms (Buckshot and KMeans), using the initial collection 20_newsgroups (n=20000). More concretely, the 50, 100 and 200 cluster centers (centroids) referred above, are generated as a result of the first phase of the algorithm, which consists of the execution of the HAC algorithm over 1000, 1415 and 2000 (consider parameter 's') initial documents, respectively. As described above (par. 5.1) the native K-Means algorithm has been observed to converge after eight iterations. The execution time and RSS measurements after the convergence of K-Means are presented, as well as the corresponding measurements for our Buckshot approach after the completion of two iterations in the second phase (for the assignment of the remaining documents). The execution time of both the algorithms was measured using 1, 3 and 10 nodes/reducers. Also, Table 8 shows the measurements of the construction of 400 clusters, generated from the execution of the HAC algorithm over 10000 initial documents (in the first phase) of the synthetic collection of 1GB (n=250000).

As it comes out from Tables 5, 6 and 7, the execution time of the Buckshot-based algorithm is significantly less than the one of K-Means in all cases. More concretely, it leads to improvements of up to 87% compared to the total time required for the convergence of K-Means. Especially when the bigger values of k are applied the time improvements are close to the above maximum for all the numbers of nodes (1, 3, 10), which implies high scalability for this approach too. Moreover, the quality of clustering is kept in acceptable levels (even better than the ones observed for the BKC algorithm) since the achieved RSS value is quite close to the final RSS value of K-Means (having a difference between only 3.5% and 5.5% in all cases).

Similar results are also given by the measurements taken for the synthetic collection of 1 GB. As it is shown in Table 8 an improvement of almost up to 80% can be observed. Additionally, in Table 8 the corresponding measurements with respect to the implementation of the proposed Buckshot-based approach in the Spark framework are also given. As it was expected (due to the in-memory iterative processing taking place in Spark) substantial additional improvements have been observed for this approach too, leading to the capability of clustering 1GB of text data in less than 10 minutes using the proposed Buckshot algorithm's adaptation. The corresponding improvement comparing to KMeans remains also to notably high levels, i.e. up to 72%.

**Table 5. Buckshot - 20_ngroups (n=20000), k=50, s=1000**

| Nodes | Execution Time | | RSS | |
|---|---|---|---|---|
| | KMeans | Buckshot | KMeans | Buckshot |
| 1 | 77m57s | 9m37s | 140.39 | 145.55 |
| 3 | 43m25s | 5m27s | 140.12 | 145.45 |
| 10 | 11m41s | 2m58s | 140.78 | 147.19 |

**Table 6. Buckshot - 20_ngroups (n=20000), k=100, s=1415**

| Nodes | Execution Time | | RSS | |
|---|---|---|---|---|
| | KMeans | Buckshot | KMeans | Buckshot |
| 1 | 81m55s | 12m18s | 135.82 | 141.97 |
| 3 | 54m30s | 7m13s | 135.19 | 142.39 |
| 10 | 15m55s | 3m43s | 136.21 | 142.36 |

**Table 7. Buckshot - 20_ngroups (n=20000), k=200, s=2000**

| Nodes | Execution Time | | RSS | |
|---|---|---|---|---|
| | KMeans | Buckshot | KMeans | Buckshot |
| 1 | 158m21s | 19min14s | 129.17 | 135.19 |
| 3 | 91m33s | 12min06s | 128.92 | 135.00 |
| 10 | 28m45s | 3min57s | 129.53 | 136.55 |

**Table 8. Buckshot - 1GB (n=250000), k=400, s=10000**

| Nodes | Execution Time | | RSS | |
|---|---|---|---|---|
| | KMeans | Buckshot | KMeans | Buckshot |
| 10 (MR) | 324m41s | 68m06s | 1131.30 | 1171.45 |
| 10 (Spark) | 33m35s | 9m34s | 1118.30 | 1163.72 |

## 5.3 Results Summarization and Discussion

As it is shown in the results of the previous paragraphs, both the proposed alternatives may improve substantially the execution time needed for clustering large scale document collections. Moreover they can be efficiently parallelized with use of Hadoop/MR and Spark frameworks, and they are scalable enough to serve as reliably efficient solutions when big text data are to be clustered. Furthermore, between the two, the Buckshot approach is experimentally proved to exhibit marginally better performance than the BKC approach, considering both the executing times and the RSS values. Consequently, in the following we pay some more attention on this approach, we summarize adequately the results drawn by its experimental evaluation and we further discuss the corresponding achievements.

More concretely, in Table 9 the total improvements in the execution time of the Buckshot-based implementation are summarized for all cases (varying number nodes and clusters) together with the RSS loss in each case. Note here that due to the random sampling employed in the Buckshot algorithm, the RSS values are not the same for different runs with the same value of parameter 'k' (number of clusters). However, the corresponding differences are quite restricted, leading to a sufficiently stable approach. Also one can observe that the RSS loss does not exceed 5.5% in any case which is a very satisfactory achievement. Also, by staring in Table IX one can observe more clearly that the improvements in the execution time are steadily high in all cases, ranging from 74.6% to 87.8% in all cases.

Moreover, the value of 71.5% in the improvement of the execution time in the case of Spark implementation over the 1GB collection, is also very satisfactory due to the fact that the KMeans algorithm is a-priori expected to gain more from its implementation in Spark framework because of it has more iterations. In other words, the absence of repetitive disk accesses flattens/eliminates the corresponding iteration overheads in both Spark implementations. A more representative view of the execution time improvements over KMeans is given in Fig. 3.

**Table 9. Execution time improvements and RSS loss**

| Nodes | Time Improvement (%) | RSS loss (%) |
|---|---|---|
| For k = 50 Clusters | | |
| 1 | 87.6 | 3.68 |
| 3 | 87.4 | 3.81 |
| 10 | 74.6 | 4.55 |
| For k = 100 Clusters | | |
| 1 | 84.9 | 4.53 |
| 3 | 86.7 | 5.33 |
| 10 | 76.6 | 4.52 |
| For k = 200 Clusters | | |
| 1 | 87.8 | 4.65 |
| 3 | 86.7 | 4.72 |
| 10 | 86.2 | 5.42 |
| For k = 400 Clusters | | |
| 10 (MR) | 79.1 | 3.55 |
| 10 (Spark) | 71.5 | 4.06 |

Finally, in Table 10 the speed-up values achieved in each case (with 3 and 10 working nodes), using both the native KMeans and our Buckshot-based approach are summarized. As it can be seen, the speed-up values achieved are quite satisfactory for the Buckshot-based approach (for 3 nodes the speedup raises up to almost 2, whereas for 10 nodes the speedup raises up to almost 5) although the hierarchical agglomerative clustering procedure is a hardly parallelizable one. It's also quite promising the fact that for 10 nodes, as the number of clusters increase the speed-up increases too, thus leading to a quite scalable behavior suitable for large scale document collections / big text data. Analogous speed-up values have been observed for the BigKClustering parallel implementation too (as it can be extracted by the values of tables 1-4). A more clear and representative view of the speed-up achievements can be found in Fig. 4.

**Table 10. Speed-up values for varying # of nodes/clusters**

| Clusters | Speed-up Values | | | |
|---|---|---|---|---|
| | KMeans (3) | Buckshot (3) | KMeans (10) | Buckshot (10) |
| 50 | 1.79 | 1.76 | 6.66 | 3.24 |
| 100 | 1.50 | 1.70 | 5.14 | 3.30 |
| 200 | 1.73 | 1.59 | 5.50 | 4.87 |

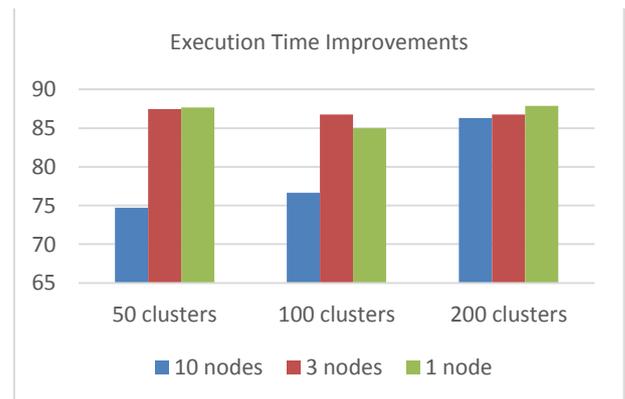

**Fig 3: Time improvements for varying # of nodes/clusters**

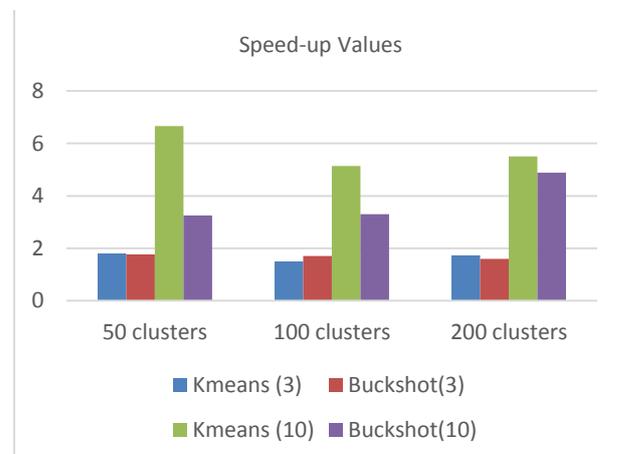

**Fig 4: Speed-up achieved for varying # of nodes/clusters**

# 6. CONCLUSION

An overview of the existing techniques for document clustering over big document collections (focusing in the use of massive distributed processing frameworks like Hadoop/MR and Spark) has been firstly presented in this paper. Next, a variation of an existing K-Means-based fast clustering technique (known as BigKClustering) is presented so that it can be applied in document clustering. Also, towards the direction of the efficient handling of big text data, a hybrid clustering approach based on a customized version of the Buckshot algorithm is presented and analyzed, which first applies a hierarchical clustering algorithm on a sample of the input dataset and then uses the results as initial centers for the assignment of the rest of the documents, with a quite restricted number of iterations. The suitable adaptations of both the proposed algorithms for big data have been suitably implemented over Hadoop/MR and Spark to efficiently handle very large text collections; and also they've been extensively tested over a real cluster environment and real text data of size up to 1GB. Moreover, as it is shown in the corresponding experiments, they achieve acceptable clustering quality as well as significant time improvements, when they are compared to the native K-Means algorithm. As a result they can be definitely considered as two very promising alternatives for clustering of big document collections.